\documentstyle[12pt]{article}

\newcommand{\underset}[2]{\sum_{#1}}
\begin{document}



\begin{center}
{\Large\bf Gravitational Wave Emission and Mass Extraction from a
Perturbed Schwarzschild Black Hole}

 \ \\[1mm]

{\sl Marcelo Costa de Lima} and {\sl Ivano Dami\~ao Soares}\\[1mm]
Centro Brasileiro de Pesquisas F\'{\i}sicas - CBPF\\[-1mm]
Rua Dr. Xavier Sigaud, 150\\[-1mm]
22290-180 -- Rio de Janeiro, RJ -- Brazil\\[-1mm]
e-mail:mclima$@@$lca1.drp.cbpf.br\\[-1mm]
ivano$@@$lca1.drp.cbpf.br\\[1cm]
{\sc Abstract}
\end{center}
A relativistic model for the emission of gravitational waves from an
initially unperturbed Schwarzschild black hole, or spherical
collapsing configuration, is completely integrated. The model
consists basically of gravitational perturbations of the
Robinson-Trautman type on the Schwarzschild spacetime. In our scheme
of perturbation, gravitational waves may extract mass from the collapsing
configuration, and the amount of mass extracted depends on the
particular gravitational wave $l$-poles emitted  $(l\geq 2)$. The
formulation allows us to define a mass emissity function
$\left.e_l=\displaystyle{\displaystyle{\frac{3b_0}{2}}}\right/
\left[\displaystyle{\frac{l(l+1)}{2}}
\left(\displaystyle{\frac{l(l+1)}{2}}-1\right)\right]$, which measures the ratio of mass
taken out from the source by a $l$-pole ($b_0$ is a small positive
parameter, of the order of the perturbations, characteristic of the
mechanism of mass emission in a Schwarzschild background spacetime).
The quadrupole emission mode is the dominant mechanism in extracting
mass from the configuration. Robinson-Trautmann perturbations also
include another mode of emission of mass, which we denote shell
emission mode: in the equatorial plane of the configuration, a
timelike $(1+2)$ shell of matter may be present, whose stress-energy
tensor is modelled by neutrinos and strings emitted radially on the
shell; no gravitational waves are present in this mode. The invariant
characterization of gravitational wave perturbations and of the
gravitational wave zone is made through the analysis of the structure
of the curvature tensor and the use of the Peeling Theorem. The time
behaviour of a $l$-pole gravitational wave has the form $\exp [-(u/3M)\
l(l+1)/2\ (l(l+1)/2-1)]$, where $M$ is the initial mass of the
collapsing configuration and $u$ may be interpreted asymptotically as
the retarded time. It follows that, for a Galilean observer at
infinity, the horizon of the final black hole configuration takes an
infinite time to form.


\newpage
\section{Introduction}

\paragraph*{}
In the study of the dynamics of formation of black holes, the final
state of the collapsing configuration is fixed by the
so-called Wheeler's lemma ``A black hole has no hair'' \cite{um}. For a
Schwarzschild black hole this lemma is sustained by the complete
analysis of scalar, electromagnetic and gravitational perturbations on
the background geometry realized in Refs. \cite{dois,sete} and
extended for non-classical fields in Refs. \cite{oito,dez}. The
analysis carried out in these references treat the perturbations as
test fields. However, in the particular case of gravitational
perturbations, the approach does not consider two important issues:
(i) how do the gravitational perturbations of radiative character
extract mass, and what is the amount of mass carried out by a
particular gravitational radiation pole emitted;  (ii) what is the
fate of the compact event horizon due to the presence of
gravitational perturbations. In this paper, our aim is to examine
these issues by considering a simple class of gravitational
perturbations of the radiative type. Our approach here will be based
on perturbations of the Robinson-Trautman type made on a
Schwarzschild background. This class of perturbations may be
understood as belonging to the family of Robinson-Trautman metrics
\cite{onze}, having the Schwarzschild geometry as a particular limit,
the advantage of which is that the definition of a mass function
appears naturally. A further advantage in the use Robinson-Trautman
perturbations is that a gravitational wave zone is well
characterized in the perturbed spacetime, such that it constitutes
a simple and suitable model for the exam of the issues discussed above. Also,
the Robinson-Trautman perturbations, if by one hand they contain just
a particular set of even-parity perturbations (in the terminology of 
Regge-Wheeler), by the other hand they are more general in the sense that
they contain a coordinate gauge dependent piece which allows us to
obtain information on how gravitational waves extract mass from the
collapsing configuration (indeed, invariant information). 

The paper is organized as follows. In Section 2, the
Robinson-Trautman metrics are introduced together with the
corresponding vacuum Einstein equations, and after particularized as
perturbations of the Schwarzschild metric. In Section 3 we examine
the structure of the curvature tensor for the Robinson-Trautman
perturbations, in order to obtain under what conditions they are true
gravitational wave perturbations and have a well defined gravitational
wave zone. The use of the Peeling Theorem is made in this
characterization. In Section 4, the integration of field equations is
completely realized, based on pertinent initial conditions, and
physical results discussed. In Section 5, the temporal gauge group
for Robinson-Trautman metrics is introduced, and an invariant mass
function defined, allowing a characterization of  mass loss from the
configuration by emission of gravitational waves. Finally in Section
6, a summing up of the resulting scenario is made.

Throughout the paper, the units are such that $c=8\pi G=1$.

\section{The Geometry and the Field Equations}

\paragraph*{}
We start by considering the family of Robinson-Trautman spacetimes
\cite{onze}, with metric given by
\begin{equation}
ds^2=A^2(u,r,\theta )du^2+2du\ dr-r^2K^2(u,\theta
)(d\theta^2+\sin^2\theta d\phi^2)\ .
\end{equation}
This geometry is non-stationary and axially symmetric, admitting the obvious
Killing vector $\partial /\partial \phi$. The components $G_{22}=0$
and $G_{33}=0$ of Einstein's equations in vacuum, together with
$G_{02}=0=G_{12}$, gives that 
\begin{equation}
A^2(u,r,\theta )=L(u,\theta )+B(u)/r+2rK'(u,\theta )/K(u,\theta )\ ,
\end{equation}
with $L$ and $K$ arbitrary functions of $u$ and $\theta$, and $B$ an
arbitrary function of $u$. Here a prime denotes $\partial /\partial
u$. The remaining vacuum equations $G_{00}=G_{01}=G_{11}=0$ lead us to
choose the function $L(u,\theta )$ as
\begin{equation}
L(u,\theta )=1/K^2-K_{\theta \theta}/K^3+K_{\theta}^2/K^4-K_{\theta} \mbox{cotg}\
\theta /K^3\ ,
\end{equation} 
where a subscript $\theta$ denotes now $\partial /\partial \theta$, resulting 
\begin{equation}
3B(u)K'/K+B'(u)+\frac{1}{2K^2\sin \theta}(L_{\theta}\sin \theta )_{\theta} =0\ .
\end{equation}

Equations (3) and (4) are the basis of our analysis in this paper.

It is easy to see that 
\begin{equation}
L=1=K\ \ \ , \ \ \ B=-2M=\mbox{const.}
\end{equation}
is a solution of (3) and (4), corresponding to the Schwarzschild
metric in outgoing Eddington-Finkelstein coordinates \cite{um}. These
coordinates are most convenient for our analysis of a non-spherical,
axially symmetric collapse with emission of gravitational waves. We
remark here that the Eddington-Finkelstein retarded coordinate $u$
may be interpreted as the Newtonian time of an observer a rest at
infinity $(r\rightarrow \infty )$ \cite{doze,treze}, and $r$ is the
parameter distance along the {\it outgoing} null geodesics
determined by the vector field $\partial /\partial r$. Of course, our
description is valid only outside the apparent horizon defined by
$A^2(u,r,\theta )=0$.

Let us now introduce Robinson-Trautman perturbations on the
Schwarzschild solution (5), namely, we take a geometry of the form
(1), (2), (3) given by
\begin{eqnarray}
ds^2 &=& \left[1+\varepsilon W(u,\theta )+
\displaystyle{\frac{-2M+\varepsilon Z(u)}{r}}+2\varepsilon
r\partial /\partial u\ Y(u,\theta )\right]du^2\nonumber \\
&+& 2dudr-r^2[1+\varepsilon Y(u,\theta )]^2(d\theta ^2+\sin^2\theta d\phi^2)\ ,
\end{eqnarray}
where $\varepsilon$ is a small parameter. As we will show later, the
curvature tensor calculated from (6) guarantees that, in general,
these perturbations do not result from mere coordinate
transformations but they are physical in the sense that the
invariants constructed with the curvature tensor calculated from (6)
are distinct from the ones of the curvature of the Schwarzschild
solution (5).

The components of the metric perturbations $h_{ab}(a,b=0,1,2,3)$ may
be expressed 
\[
h_{ab}=\left(
\begin{array}{ccccc}
W(u,\theta)+Z(u)/r+2r & \partial_uY(u,\theta ) & 0 & 0 & 0\\
0 & & 0 & 0 & 0 \\
0 & & 0  & 2r^2Y(u,\theta ) & 0\\
0 & & 0 & 0 & 2r^2Y(u,\theta  )\sin^2\theta \\
\end{array}
\right)\ \ (6.a)
\]
After the integration of the field equations (3) and (4) for the
perturbed metric, when the functions $W(u,\theta )$ and $Y(u,\theta )$
are determined,
it will result that perturbations (6.a) are actually split into two
parcels, only one of which will have angular dependence. This latter
parcel is a particular type of even-parity perturbations, in the
terminology of Regge and Wheeler (Ref. \cite{dois}), or of polar-type
perturbations in the 
terminology of Chandrasekhar (Ref. \cite{quatorze}), and are already
in the canonical form for even-parity perturbations, as given in Ref.
\cite{dois}. The other parcel will contain, as its entries, the
function $Z(u)/r$ and separation constants appearing in the
integration. We might be tempted to interpret
$B(u)=-2M+\varepsilon Z(u)$ as a time-dependent mass term, or $Z(u)$
as a perturbation in the mass $M$ of the Schwarzschild background.
However, as will be discussed in Section 5, we can always choose a
temporal gauge in which $B(u)$ is a constant. Therefore it is not
possible to attach a meaning to the time dependence of $B$, unless
the time coordinate $u$ may be fixed in an independent way, for
instance, as the retarded time coordinate of the Schwarzschild
background. This is a crucial point in our paper, which shall be dealt
with in Section 5. There we will show  that we can define an invariant mass
function which coincides with $B(u)$ for large values of $u$.

Inserting (6) in the field equations (3) and (4), we obtain in first
order in $\varepsilon$,
\begin{equation}
dZ/du-6M\ \partial Y/\partial u+ \frac{1}{2 \sin \theta} [(\sin \theta \ 
W_{\theta})_{\theta}]=0\ ,
\end{equation}
\begin{equation}
Y_{\theta \theta}+Y_{\theta}\mbox{cotg}\ \theta +2Y=-W\ .
\end{equation}
The form of the above equations suggests the following separation Ansatz
\begin{eqnarray}
Y(u,\theta  ) &=& y(\theta )N(u)\ ,\\
W(u,\theta ) &=& w(\theta )N(u)\ , \nonumber
\end{eqnarray}
yielding from (7) and (8), 
%
\begin{eqnarray}
&&-\displaystyle{\frac{6M}{N}}\ [dN/du]=a_0=\mbox{const}.\ ,\\
&&\displaystyle{\frac{1}{N}}\ [dZ/du]=b_0=\mbox{const}.\ , 
\end{eqnarray}
\begin{eqnarray}
w_{\theta \theta} &+& w_{\theta\ }\mbox{cotg}\ \theta +2a_0y+2b_0=0\ , \\
y_{\theta \theta} &+& y_{\theta}\ \mbox{cotg}\ \theta +2y=-w\ .
\end{eqnarray}
%
Equations (10) and (11) determine uniquely (up to a temporal gauge
freedom to be discussed later) the Robinson-Trautman gravitational
perturbations of Schwarzschild black hole of mass $M$.

Before proceeding into the task of integrating equations (10) and
(11), taking into account the initial conditions connected to the
emission of gravitational waves, we must discuss the physical nature
of the perturbations considered and what are the new features present
in the spacetime described by (6). In this way, we examine, in the
next section, the structure of the Weyl tensor of (6).

\section{The Structure of the Weyl Tensor and the Gravitational Wave Zone}

\paragraph*{}
Let us introduce the semi-null tetrad basis determined by the 1-forms 
\begin{eqnarray}
O^0 &=& du\ , \nonumber \\
O^1 &=& A^2/2\ du+dr , \\ 
O^2 &=& r\ K \ d\theta \ , \nonumber \\
O^3 &=& r\ K \ \sin \theta \ d\phi \nonumber \ ,
\end{eqnarray}
where the metric (1) assumes the from $ds^2=g_{AB}O^AO^B$, with 
\[
g_{AB}=\left(
\begin{array}{cccc}
0 & 1 & 0 & 0 \\
1 & 0 & 0 & 0\\
0 & 0 & -1 & 0\\
0 & 0 & 0 & -1\\
\end{array}
\right)\ \ .
\]
The tetrad basis has the physical property that it is parallely
propagated along the null geodesics determined by $\partial /\partial
r$. As we shall see, these {\it outgoing} null geodesics are the
propagation direction of the gravitational waves. In this basis, the
non-zero Weyl tensor components are given by
\begin{eqnarray}
C_{2323} &=& -C_{0101}=2\ C_{0212}=B(u)/r^3=[-2M+Z(u)]2/r^3\ , \\
C_{0303} &=& -C_{0202}=A(u,\theta )/r^2\ \ \ , \ \ \ 
C_{0303}=L_{\theta}/2Kr^2\ , \\
C_{0202} &=& -C_{0303}=-D(u,\theta )/r\ , 
\end{eqnarray}
where the functions $A$ and $D$ are
\begin{eqnarray}
A(u,\theta ) &=& \displaystyle{\frac{1}{4K^2}}\ (-L_{\theta \theta}+2L_{\theta}K_{\theta}/K+
L_{\theta}\mbox{cotg}\ \theta ) \ ,\\
D(u,\theta ) &=& \displaystyle{\frac{1}{2K^2}}\ \partial_u[(K_{\theta}/K)_{\theta}-
K_{\theta}/K\mbox{cotg}\ \theta -(K_{\theta}/K)^2]\ .\nonumber 
\end{eqnarray}
We note that the $r$-dependence of the components (13.a,b,c) is ,
respectively, $1/r^3,\ 1/r^2$ and $1/r$. Indeed, from (13), we may
express 
\begin{equation}
C_{ABCD}=D_{ABCD}/r^3+III_{ABCD}/r^2+N_{ABCD}/r\ ,
\end{equation} 
where $D_{ABCD},\ III_{ABCD}$ and $N_{ABCD}$ are of algebraic type
$D$, type $III$ and type $N$ in the Petrov classification
\cite{quinze,dezeseis,dezesete}, respectively, and have the vector
field $k=\partial /\partial r$ as a principal null direction (cf.
(17) below). In the coordinate basis, they have the property of
being covariantly constant along the null direction $k$ \cite{onze}.
For the Schwarzschild geometry, only type $D$ terms are present, and
the non-zero components of the Weyl tensor are 
\begin{equation}
C_{2323}=-C_{0101}=2\ C_{0212}=-4M/r^3\ .
\end{equation}
Comparing (13) with (16) we can see that, indeed, the metric (6) is
a true perturbation of the Schwarzschild geometry.

From (15) we can now establish an invariant characterization of the
presence of gravitational waves in the perturbed spacetime (6), and
of a corresponding gravitational radiation wave zone. This is based on two
pillars: 
\begin{description}
\item
[(i)] the Peeling Theorem (for the linearized Riemann tensor of
retarted multipole fields, see Refs. \cite{vum,vdois}; for the
general case, see Ref. \cite{vtres}; for a review, including peeling
properties of the Maxwell tensor, see Ref. \cite{vinte});
\item
[(ii)] the analysis of the spacetime of gravitational wave solutions
of Einstein's equations, and their relation to electromagnetic waves
in Maxwell's theory \cite{um,onze,dezoito,dezenove,vinte}.
\end{description}

The Peeling Theorem states that the Weyl tensor (or vacuum Riemann
tensor) of a radiative gravitational bounded source, expanded in
powers of $(1/r)$, has the general form
\begin{equation}
C_{ABCD}=N_{ABCD}/r+III_{ABCD}/r^2+II_{ABDC}/r^3+I_{ABCD}/r^2+\cdots
\end{equation}
where $r$ is the parameter distance defined along the null geodesics
determined by the null vector field $k=\partial /\partial r$. The
quantitites $N_{ABCD},\ III_{ABCD},\ II_{ABCD}$ and $I_{ABCD}$, when
expressed in the coordinate basis, have vanishing covariant
derivatives along the null vector field $k$. They are of the
algebraic type $N$, $III$, $II$ and $I$, respectively, in the Petrov
classification. The direction of propagation $k$ is a repeated
principal null direction \cite{dezeseis,dezesete}, of the Weyl
tensor to order $r^{-4}$, and satisfies 
\begin{eqnarray}
N_{ABCD}k^D &=& 0\ , \nonumber \\
III_{ABC[D}k^Ck_{E]} &=& 0\ , \nonumber \\
II_{ABC[D}k_{E]}k^Bk^C &=& 0 \ ,\\
k_{[E}I_{A]BC[D}k_{F]}k^Bk^C &=& 0 \ .\nonumber
\end{eqnarray}
If the spacetime is such that $N_{ABCD}$ is non-zero, then,
for large values of the distance parameter, the curvature tensor has
the approximate asymptotic expression $C_{ABCD}=N_{ABCD}/r$, that is,
it is of Petrov type $N$, with the degenerate principal null
direction given by $k$. This is the curvature tensor of a
gravitational wave spacetime \cite{um,onze,dezoito,dezenove,vinte},
with propagation vector $k$ (cf. (18)). In other words, the field
looks like a plane wave at large distances. The non-vanishing of the
scalars $N_{ABCD}$ is then taken as a invariant criterion for the
presence of gravitational waves, and the asymptotic region (where the
$0(1/r)$ term in (17) is dominant) defined as the wave zone.

Now if we compare (15) with (17), we see that the invariant
condition for gravitational wave perturbations in (6) is that (cf. (14)) 
\begin{equation}
D(u,\theta )=-\varepsilon /2 [y_{\theta \theta}
-y_{\theta}\mbox{cotg}\ \theta ]dN/du=-
\varepsilon (a_0/12M)[y_{\theta \theta}-y_{\theta}\mbox{cotg}\ \theta
]N\neq 0 \ .
\end{equation}

From the comparison of (13) and (15) with (17), we can see that the
perturbations (6) are not general, in the algebraic sense. In fact,
(6) is algebraically special \cite{vquatro}. However, the credo in
the literature is that exact radiation fields must be algebraically
general. This comes from the analysis of the linearized vacuum
Riemann tensor of a retarded multipole field \cite{vinte,vdois}, the
form of which is the same as (17), and the consideration that exact
radiation fields are at least as complicated as their linearized
approximation. For instance, in the linear approximation, a static 
quadrupole gives rise to terms proportional to $r^{-5}$, while terms
going as $r^{-1},\ r^{-2},\ r^{-3}$ and $r^{-4}$ arise if the
quadrupole moment varies in time \cite{vum}. Nevertheless the problem
of the structure of the source of a radiation field in the full
non-linear theory is still open (we should mention a tentative
definition of multipole structure of the gravitational source, based
upon a detailed analysis of the linearized theory \cite{vcinco}).
From the point of view of Einstein's equations, the expression (15)
may well be sustained while it is inconsistent from the point  of
view of the linearized theory. We assume that (15) can well represent
a particular structure of the bounded source of the field, which
indeed radiates even-parity multipoles, as we will show. For our
purposes, these even-parity perturbations of the Schwarzschild
geometry are sufficient to give us an answer about the question of how a
collapsing star (or black hole), when perturbed, may loose mass by
emitting a particular gravitational radiation $l$-pole.

\section{The Integration of Field Equations and Initial Conditions}

\paragraph*{}
Equation (10) can be immediately integrated to
\begin{eqnarray}
N &=& N_0\exp [(-a_0/6M)u]\ , \\
Z &=& Z_0-(6b_0M/a_0)N_0\exp  (-a_0u/6M)\ ,\nonumber 
\end{eqnarray} 
where $N_0$ and $Z_0$ are integration constants. The model envisaged
for the initial conditions which fix the integration constants $Z_0$
and $N_0$ is that of an initially collapsing spherical star, or
unperturbed Schwarzschild black hole of mass $M$, which, in a given
time, say $u=0$, starts to emit gravitational radiation in the
Robinson-Trautman regime, that is, in accordance to (6). In view of
this, gravitational waves are emitted with {\it initial amplitude}
(this is a provisional denomination, justified in Refs.
\cite{dezoito,dezenove,vinte}) $N_0$ and we must have $Z(u=0)=0$ (cf.
Ref. \cite{vseis}). The mass perturbation (21) is then fixed to
\begin{equation}
Z=(6b_0M/a_0)[1-N_0\exp (-a_0u/6M)]\ .
\end{equation}
The gravitational waves emitted carry the information that the system was
switched on in $u=0$, through the finite discontinuity of the
$O(1/r)$ components of the Riemann tensor in the first gravitational
wave front $u=0$ emitted. This discontinuity is determined by
\begin{equation}
[N_{ABCD}](u=0)=-\varepsilon a_0N_0/6M\ , 
\end{equation}
up to an angular factor (cf. (13c)). We remark that the discontinuity
appearing in (6), due to the time derivative
$\displaystyle{\frac{dN}{du}}(u=0)$, can be 
properly eliminated by a coordinate transformation, but its presence
in the scalars (13) of the Weyl curvature tensor is unavoidable. 

In the integration of  the angular equations (11) we must distinguish two cases:\\
(i) $a_0\neq 0$; the requirement of solutions regular at $p=0$ and
$p=\pi$ demands that
\begin{eqnarray}
a_0 &=& : A_l=2\ \frac{l(l+1)}{2}\left[ \frac{l(l+1)}{2}-1\right]
\end{eqnarray}
\vspace*{-1cm}
\begin{eqnarray}
w_l &=& 2\ b_0/A_l+w_{0l}P_l(\cos \theta )\\
y_l &=& -b_0/A_l+\left[\frac{l(l+1)}{2A_l}\right]w_{0l}P_l(\cos \theta )\ ,
\end{eqnarray}
$\!\!\!$where $l$ is a non-negative integer and $w_{0l}$ is an arbitrary
non-zero constant \cite{vsete}. Here $P_l(\cos \theta )$ is the
Legendre polynomial with angular momentum $l$. The condition $a_0=: A_l\neq 0$ defining
case (i) implies that 
\begin{equation}
l\geq 2\ ,
\end{equation}
that is, only quadrupole or higher-order poles gravitational
radiation fields are emitted, as should be expected. Obviously
outgoing gravitational waves are present in the spacetime: condition
(19) for holds $a_0\neq 0$, with wave zone defined by the $O(1/r)$
non-zero components of the Weyl tensor
\begin{equation}
C_{0202}=-C_{0303}=-\frac{\varepsilon}{r}\ 
\frac{l(l+1)w_{0l}}{24M}\ (-2\cos 
\theta dP_l/d\theta +l(l+1)P_l)N_0\exp (-A_lu/6M)\ .
\end{equation}
In sum, the general Robinson-Trautmann perturbations $h_{ab}$ for
this case can be split into
\begin{equation}
h_{ab}=h^{(1)}_{~~ab}+h^{(2)}_{~~ab}
\end{equation}
where
\begin{equation}
h^{(1)}_{~~ab}=\left(
\begin{array}{lccc}
A_l/(l(l+1))-(A_l/6M)r & 0 & 0 & 0\\
\ \ \ \ \ \ \ \ \ \ 0 & 0 & 0 & 0\\
\ \ \ \ \ \ \ \ \ \ 0 & 0 & r^2 & 0\\
\ \ \ \ \ \ \ \ \ \ 0 & 0 & 0 & r^2\sin^2\theta\\
\end{array}
\right)\cdot 
\end{equation}
\[
\cdot \ \frac{l(l+1)}{A_l}w_{0l}P_l(\cos \theta ) N_{0l}\exp (-A_lu/6M)\ , 
\]
and
\begin{eqnarray}
h^{(2)}_{~~ab} &=& \displaystyle{\frac{Z(u)}{r}}+
\left(\displaystyle{\frac{2}{A_l}}+\frac{r}{3M}\right)b_0N_{0l}\exp 
[(-A_l/6M)u])\delta_a^0\delta_b^0 +\nonumber \\
&-& \displaystyle{\frac{2r^2}{A_l}}b_0N_{0l}\exp [(-A_l/6M)u]
(\delta_a^2\delta^2_a+\sin^2\theta \delta^3_a\delta^3_b)\ ,
\end{eqnarray}
with $Z(u)$ given by (21). The perturbations (28) are already in the
canonical form for even parity perturbations, in the terminology of
Regge and Wheeler \cite{dois}. Under the inversion operation
$\theta \rightarrow \pi -\theta$, they transform as
$h^{(1)}_{~~ab}\rightarrow (-1)^lh^{(1)}_{~~ab}$. Perturbations (29)
are denoted ``mass perturbations''. Although dependent on the
temporal gauge, they will cause an effective decrease of the mass of
the system.  This analysis will be done in Section 5, where an {\it
invariant} mass perturbation function (due to outgoing gravitational
waves) is defined. The final
configuration will be a Schwarzschild geometry with mass smaller than
the original mass, the effective decrease of mass being dependent on
the particular pole emitted.

The non-radiatable modes correspond to the case\\
\noindent
(ii) $a_0=0:$ as expected, no gravitational radiation is present
($N=$ const, $C_{0202}=-C_{0303}=0$, cf. (19) and (20)).  Solutions of 
equations (11) for $a_0=0$ and $b_0\neq 0$ are, in general, singular
at $p=0$ and and $p=\pi$. We select
the particular set
\begin{eqnarray}
w &=& w_0+2b_0\ ln(1+\cos \theta )\ , \nonumber \\
y &=& -(w_0+b_0)/2-b_0\ ln(1+\cos \theta )+y_0P_1(\cos \theta )\ ,
\end{eqnarray}
which are regular at $\theta =0$, and
\begin{eqnarray}
w &=& w_0+2b_0\ ln(1-\cos \theta )\ ,\nonumber \\
y &=& -(w_0+b_0)/2-b_0\ ln(1-\cos \theta )+y_0P_1(\cos \theta ) ,
\end{eqnarray}
which are regular at $\theta =\pi$. We remark that the parcel
$y_0P_1(\cos \theta )$ appearing in both Eqs. (30) can be eliminated by a
convenient coordinate transformation, which asymptotically is
interpreted as an infinitesimal Lorentz boost with velocity parameter
$\varepsilon y_0$. Solutions (30) coincide at $p=\pi /2$. We
therefore define the continuous solution in $[0,\pi ]$ as the union
of (30.a) in $[0,\pi /2]$, and (30.b) in $[\pi /2,\ \pi ]$. Although
continuous in $p=\pi /2$, its first derivative has a finite
discontinuity, defining physically a shell of matter at the equatorial
plane. This shell can be modelled by neutrinos and strings being
emitted radially in the $1+2$ spacetime of the shell. This
configuration was examined in Ref. \cite{voito}, and will be referred
to here as the shell emission mode for $b_0\neq 0$. The parameter
$b_0$ is proportional to the neutrino flux emitted on the shell;
this, in fact, gives us a direct mechanism for the measurement of
$b_0$. The mass function
pertubation of this mode is 
\begin{equation}
Z(u)=b_0N_0u+\mbox{const}.
\end{equation}

We interpret $b_0$ as the mass emissivity parameter of the collapsing
configuration, either in gravitational wave emission modes $l\geq 2$, or in a
shell emission mode $a_0=0$. This parameter appears to be
characteristic of mass variation in the Schwarzschild background, and
we conjecture that its existence is independent of the particular
perturbations considered. In the next Section, this parameter will
allow us to define a mass emissivity function associated to a
$l$-pole. Physical considerations implies obviously $b_0\geq 0$. 

We must now comment on the superposition of the modes $l\geq 2$, and
also of the shell emission mode with a $l\geq 2$ mode. It is easy to
verify that the solutions 
\[
W=\underset{l}{\Sigma}C_lW_l\ \ \ , \ \ \ Y=\underset{l}{\Sigma}C_lY_l\ \ \ , \ \ \ 
Z=\underset{l}{\Sigma}C_lZ_l\ ,
\]
where $W_l,\ Y_l$ and $Z_l$ are given by
\begin{equation}
\begin{array}{rcl}
\left(
\begin{array}{c}
W_l\\
Y_l
\end{array}
\right)
&=& \left(
\begin{array}{l}
2b_0/A_l+w_{0l}P_l(\cos \theta )\\
-b_0/A_l+[l(l+1)/2A_l]w_{0l}P_l(\cos \theta )\\
\end{array}
\right)\cdot N_0\exp (-A_lu/6M)\, ,\\ \
\\ 
Z_l &=& (6b_0M/A_l)N_{0l}[1-\exp (-A_lu/6M)] \, , 
\end{array}
\end{equation}
satisfy the field equations (7) and (8). Note that necessarily the same
coeficients $C_l$ appear in the three linear combinations. Also a
linear combination of a set solutions (32) with a solution for the
shell emission mode satisfy (7) and (8).

\section{The Temporal Gauge Group, the Invariant Mass Function and
The Mass Emissity}

\paragraph*{}
The Robinson-Trautmann metrics described by (1), (2), as well the
field  equations (3) and (4), are invariant under a subgroup of
coordinate transformations  $(u,r,\theta ,\phi )\rightarrow (\bar{u},\bar{r},\theta ,\phi
)$ described by
\begin{eqnarray}
\bar{r}&=&rF\ , \nonumber \\
d\bar{u} &=& du/F\ ,
\end{eqnarray}
where $F$ is an arbitrary function of $u$. Under  (33.a), the
quantities appearing in the metric (1)--(2) transform as
\begin{eqnarray}
\bar{L} &=& LF^2\ , \nonumber \\
\bar{B} &=& BF^3\ , \\
\bar{K} &=& K/F\ ,\nonumber 
\end{eqnarray}
In other words, under (33) it is enough to replace, into Equations
(1) to (4), unbarred coordinates and variables by the corresponding
barred ones. We remark the generality of (33), because they are 
general transformations leaving the Weyl scalars (13) invariant.

From (33.b) we can see that the time dependence of the mass function
$B(u)$ is not an invariant, but  we define the invariant mass aspect or mass function \cite{vnove}
\begin{equation}
m(u,\theta )=-BK^3/2
\end{equation}
which, for our purposes, will provide an useful invariant
characterization of the mass variation due to the emission of
gravitational waves. It is interesting to note how the Weyl curvature
scalars (13.a), associated to the Newtonian component of the field
(cf. (16)), is expressed in terms of this invariant function. Indeed,
\[
C_{2323}=-C_{0101}=2C_{0212}=\frac{BK^3}{(rK)^3}\ ,
\]
justifying our characterization of $BK^3$ as invariant mass function.
We also note that these are the sole Weyl scalars where the mass
variation function is present. For the case of the Schwarzschild spacetime, it
follows immediately that the invariant definition (34) yields exactly
the constant mass parameter $M$. This can be checked either for the
expression (1) with (5), or for the expression of the Schwarzschild
geometry given by (6), with (20), (21) and (24), and $w_{0l}=0$ (cf.
also Ref. \cite{vsete}). 

From (6), (9), (20), (21) and (24.b) we obtain
\begin{equation}
m(u,\theta )=M-\frac{3\varepsilon MN_{0l}}{A_l}\left\{b_0-
\frac{l(l+1)}{2}\ w_{0l}P_l
(\cos \theta )\exp [(-A_l/6M)u]\right\}\ .
\end{equation}
This invariant mass function (35) will coincide with the
gauge-dependent one given from (21), for $u\rightarrow \infty$,
resulting in the constant value
\begin{equation}
m=M-3\varepsilon b_0MN_{0l}/A_l\ .
\end{equation}
Therefore (36) can be properly interpreted as the invariant mass of
the final configuration. Hence, in the limit $u\rightarrow \infty$,
the geometry will be the one of a Schwarzschild black hole with
invariant mass given by (36), smaller than the mass $M$ of the
original configuration $(u<0)$. We note that (35) can be interpreted
as mass only for some specific limits, for instance $u\rightarrow
\infty$, when no gravitational waves are present. The total amount 
of mass extracted by each gravitational wave pole emitted is given by
\begin{equation}
\Delta m=3\varepsilon b_0MN_{0l}/A_l=
\frac{3\varepsilon b_0MN_{0l}}
{2[l(l+1)/2(l(l+1)/2-1)]}\ .
\end{equation}
The $l=2$ quadrupole emission appears then as being the most
effective mode in the mechanism of extracting mass by emission of
gravitational waves. From (37) we are led to define the mass
emissivity factor $e_l$ as the mass fraction extracted by a $l$-pole
gravitational wave of unit amplitude, that is, 
\begin{equation}
e_l=:\Delta m/MN_{0l}=\frac{3\varepsilon b_0}{2[l(l+1)/2(l(l+1)/2-1)]}\ .
\end{equation}
We conjecture that the $l$-pole dependence given in (38) is a
characteristic of the \linebreak Schwarzschild black hole, and independent of
the perturbations which gave rise to the gravitational waves emitted.

The following remarks are pertinent here. As we have seen, in our
perturbative scheme for the Schwarzschild geometry, a temporal gauge
infinitesimal transformation can always be performed such that
$\bar{Z}=0$. From this it could be incorrectly inferred that the
results (36)-(37) on mass decrease may be eliminated, because the
parameter $b_0$ would then not appear in the solutions in the new
gauge. This is not the case, since the perturbations in the new gauge
are 
\begin{eqnarray}
\bar{W} &=& W+\frac{1}{3M}\ Z \nonumber \\
\bar{Y} &=& Y-\frac{1}{6M}\ Z\nonumber \\
\bar{Z} &=& 0\nonumber
\end{eqnarray}
where $Z$ is given in (21). Therefore $b_0$ is still present in the
solutions and in the limit $\bar{u}\rightarrow \infty$ when no
gravitational waves are present, the usual Schwarzschild mass
definition yields (36). Analogously we could have started the
integration directly in a temporal gauge where $Z=0$. Eqs. (7) and
(8), and the separation Ansatz $Y=y(\theta )N(u)+L(u)$,
$W=w(\theta )N(u)-2L(u)$ yields the same result (36) in the
limit $u \rightarrow \infty$. 

We must finally comment the existence of solutions with $b_0=0$ and
$a_0\neq 0$, corresponding to outgoing gravitational waves which do
not extract mass from the configuration. No gravitational wave
experiments will distinguish the cases $b_0=0$ and $b_0>0$. The only
direct experiment to measure $b_0$ is the detection of the shell
emission mode; in this case, $b_0$ is proportional to the flux of 
neutrinos emitted radially on the shell. If we adhere to the general
formulation of Bondi, Van der Burg and Metzner \cite{vnove} for the
emission of gravitational waves by an axially symmetric bounded source, we
should then discard the $b_0=0$ solutions as being physically not
meaningful. Their presence is possibly due to the simple form we have
assumed to the Robinson-Trautmann perturbations. We intend to return to
this point in the future. 

\section{Conclusions and Final Remarks}

\paragraph*{}
In this paper we discussed gravitational wave perturbations of a
static black hole or a spherical collapsing star, the exterior
spacetime of which is described by the Schwarzschild metric, by using
the so-called Robinson-Trautman perturbations. This class of radiative
type perturbations belongs to the family of Robinson-Trautman
metrics, which have the Schwarzschild geometry as a particular limit.
Einstein's field equations are integrated completely. We obtain that
Robinson-Trautman perturbations are constitued of two distinct
pieces, one of them being a particular set of even-parity perturbations
(in the terminology of the formalism of Regge-Wheeler). Although the second
piece is coordinate gauge dependent, it allows us to obtain
information on how mass is extracted from the configuration by the
emission of gravitational waves. An invariant mass function is
defined, and it results that gravitational waves may extract mass from
the collapsing configuration, and the amount of mass extracted
depends on the particular gravitational wave $l$-poles emitted. We
are able to define a mass emissivity function
\[
e_l=\frac{3b_0}{\{l(l+1)[l(l+1)/2-1]\}}\ ,
\]
which measures the ratio of mass taken out from the source  by a
$l$-pole. Here $b_0$ is a small positive parameter, of the order of
the perturbations and characteristic of the mechanism of emission.
The quadrupole emission mode $(l=2)$ is the dominant mechanism in
extracting mass from the configuration. Robinson-Trautman
perturbations also includes a mechanism of extracting mass from the
collapsing configuration, denoted shell emission mode, where no
gravitational waves are present; this mode corresponds to neutrinos 
and strings emitted radially on a shell of mass at the equatorial
plane of the configuration. Although this mode might be superposed to
gravitational wave models $(l\geq 2)$, we could have used it as a
mechanism to perturb the configuration, previous to the emission of
gravitational waves. In the Robinson-Trautman regime, the presence of
gravitational waves, and of a gravitational wave zone are well
characterized by the analysis of the asymptotic structure of the Weyl
curvature tensor. From the result of the integration of Einstein's
equations, namely Eqs. (11), and from the condition $a_0\neq 0$
necessary for the presence of gravitational waves and a gravitational
wave zone (cf. Eq. (19)), only, it follows that the lowest
gravitational wave pole emitted is the quadrupole $(l=2)$. This
result was expected from classical radiation theory for a spin-2
field and from  Einstein's linearized theory for gravitational waves.

Extraction of mass appears to be an essential characteristic of
gravitational wave emission and we conjecture that the $l$-dependence
expressed in Eq. (38) is a 
characteristic of the Schwarzschild exterior geometry only, and does
not dependent on the class of perturbations which gave rise to the
gravitational waves emitted.

The time behaviour of a $l$-pole gravitational wave has the form
\[
\exp [-(u/3M)l(l+1)/2(l(l+1)/2-1]\ ,
\]
where $M$ is the initial mass of
the collapsing configuration. It follows that, for a Galilean
observer at infinity, the horizon of the final black hole
configuration takes an infinite time to form.

\section*{Acknowledgements}

\paragraph*{}
The authors are grateful to Conselho Nacional de Desenvolvimento
Cient\'{\i}fico e Tecnol\'ogico for financial support.


\begin{thebibliography}{20}
\bibitem{vseis} In this scenario we do note take into account that
the perturbation of the collapsing star or of a static black hole
does actually occur in a finite period of time: indeed, for
$u<u_0<0$, we have the Schwarzschild geometry; in the interval
$u_0<u<0$, the black hole or star is perturbed so that in $u=0$ the
configuration starts to emit gravitational waves. Whatever the
details of the geometry and of the mechanism of perturbation in the
interval $u_0<u<0$, the fact that the configuration begins to emit
gravitational waves in $u=0$, and goes on emitting according to 
(6), necessarily implies that the curvature tensor will have a finite
discontinuity in the first emitted gravitational wave front $u=0$.
This is basically independent of the perturbating mechanisms in the
interval $u_0<u<0$, and legitimates the limit $u_0\rightarrow 0$,
provided that the total mass $M$ in $u=0$ eventually includes the
contribution of the perturbing mechanism in that interval.
\bibitem{vsete} We note that the solutions with $w_{0l}=0$ correspond
to the Schwarzschild geometry expressed in another temporal gauge.
\end{thebibliography}
\end{document}